\documentclass[aip,graphicx]{revtex4-1}
\usepackage{amssymb,amsmath,amsfonts,latexsym}
\usepackage{graphicx,xcolor}
\usepackage[english]{babel}
\usepackage{tabularx}
\usepackage{mathtools}
\usepackage{mathrsfs} 
\usepackage{graphicx, texdraw}
\usepackage{lineno}
\usepackage[normalem]{ulem}

\begin{document}
%\linenumbers
\title{Quantitative investigation of quantum emitter yield in drop-casted hexagonal boron nitride nanoflakes}

\author{Tom Kretzschmar}
\affiliation{Institute of Applied Physics, Abbe Center of Photonics, Friedrich-Schiller-University, D--07745 Jena, Germany}

\author{Sebastian Ritter}
\affiliation{Institute of Applied Physics, Abbe Center of Photonics, Friedrich-Schiller-University, D--07745 Jena, Germany}

\author{Anand Kumar}
\affiliation{Institute of Applied Physics, Abbe Center of Photonics, Friedrich-Schiller-University, D--07745 Jena, Germany}
\affiliation{Department of Computer Engineering, School of Computation, Information and Technology, Technical University Munich, D--80333 Munich, Germany}

\author{Tobias Vogl}
\affiliation{Institute of Applied Physics, Abbe Center of Photonics, Friedrich-Schiller-University, D--07745 Jena, Germany}
\affiliation{Department of Computer Engineering, School of Computation, Information and Technology, Technical University Munich, D--80333 Munich, Germany}

\author{Falk Eilenberger}
\affiliation{Institute of Applied Physics, Abbe Center of Photonics, Friedrich-Schiller-University, D--07745 Jena, Germany}
\affiliation{Fraunhofer Institute for Applied Optics and Precision Engineering IOF, D--07745 Jena, Germany}
\affiliation{Max Planck School of Photonics, D--07745 Jena, Germany}

\author{Falko Schmidt}
\email{schmidtfa@ethz.ch}
\affiliation{Institute of Applied Physics, Abbe Center of Photonics, Friedrich-Schiller-University, D--07745 Jena, Germany}
\affiliation{Current address: Nanophotonics Systems Laboratory, ETH Zurich, CH--8092 Zurich, Switzerland}

\date{\today}


\begin{abstract}
Single photon emitters (SPEs) are a key component for their use as pure photon source in quantum technologies.
In this study, we investigate the generation of SPEs from drop-casted hexagonal boron nitride (hBN) nanoflakes, examining the influence of the immersion solution and the source of hBN.
We show that, depending on the utilized supplier and solution the number and quality of the emitters changes.
We perform a comprehensive optical characterization of the deposited nanoflakes to assess the quality of the generated SPEs.
We show quantitative data on SPE yields, highlighting significant variations among solvents and different sources of hBN.
This holds particular significance for employing drop-casted nanoflakes as SPE sources in quantum communication, sensing, and imaging.
Our method is easily expandable to all kinds of surfaces and can be done without requiring complex fabrication steps and equipment, thus providing the necessary scalability required for industrial quantum applications.
\end{abstract}

\maketitle

\section{Introduction}

\noindent Single photon emitters (SPEs) have recently gained great importance for their use as pure photon sources in quantum technologies.
As a result, they have become an integral component in the fields of quantum key distribution, metrology, computing, and ghost imaging \cite{aharonovich2016,zeng2022,steinlechner2013,kok2007,zerom2011}.
Two-dimensional (2D) materials have been shown to exhibit many types of defect-based SPEs. 
Among the potential 2D materials, hexagonal boron nitride (hBN) emerges as an excellent host material for SPEs due to its emission at room-temperature, long-term stability, low cost, and widespread availability \cite{Tran2016Q,tran2016,vogl2017,sajid2020}.
Current fabrication methods for SPEs from hBN typically involve manual exfoliation from pristine bulk crystals, followed by deterministic transfer for post-processing \cite{vogl2018,chen2021}. 
Active emission from exfoliated flakes has been demonstrated through processes such as thermal annealing \cite{vogl2018,chen2021}, stress-generation on top of nanopillars \cite{li2021}, or after high energy beam exposure of such flakes \cite{tran2016,Choi2016,Chejanovsky2016,Vogl2019,Gao2021,kumar2023}.
Nevertheless, these methods are limited due to labour-intensive transfer processes.
Consequently, this impedes the progress towards integrated single photon sources in conjunction with diverse technology platforms, particularly their efficient coupling into nanophotonic systems.
Several alternative methods have emerged aiming to address this challenge, including chemical vapour deposition (CVD) \cite{li2021,hausler2021}, hydrothermal reaction \cite{Vogl:2019aa,chen2021bottom}, and drop-casting \cite{chen2021Solvent,barelli2023,schauffert2022}.
These methods offer the potential for depositing flakes onto diverse substrates, irrespective of their geometric shape, and eliminate the need for precise alignment tools.
While CVD methods are rapidly advancing \cite{zhang2023}, leading to an enhancement in the quality of synthesized crystals, drop-casting utilizes nanoflakes derived through solvent mediated exfoliation from pristine bulk crystals.
The latter bypasses the requirement for complex fabrication equipment.
Drop-casting involves dispensing a droplet of a suspension of nanoflakes onto a substrate that remain on the surface after the solvent has evaporated \cite{kumar2020}.
The choice of solvent plays a critical role, influencing the accumulation and size of suspended nanoflakes, an aspect that has only recently been investigated for hBN \cite{chen2021Solvent,marsh2015,smith2021,martinez2023}.
Beyond the flake's geometry, the presence of organic solvent molecules can strongly influence SPE quality by activating additional native point defects on hBN \cite{Ronceray2023}.
Experimental studies utilizing drop-casting of hBN nanoflakes observed the generation of SPEs and deposited them onto various substrates and waveguides \cite{chen2021Solvent,schauffert2022}.
Despite that these methods are capable of producing SPEs from single hBN nanoflakes, systematic approaches to assess their yield is still in its infancy \cite{islam2023}.
While previous studies investigating the influence of solvent on SPE quality and yield have utilized exfoliated flakes \cite{Ronceray2023}, similar systematic approaches for drop-casted nanoflakes are missing \cite{islam2023}.
Efficient SPE generation and their implementation into quantum technologies thus require further development and optimization of such processes.\\
This study addresses this open challenge by systematically evaluating the impact of solvent selection for the generation of SPEs from hBN nanoflakes using the drop-casting approach.
Following a comprehensive optical characterization and analysis of photoluminescence properties of the deposited nanoflakes, we have evaluated the produced SPEs based on their brightness and purity.
Crucially, we report quantitative data on SPE yield and emitter quality, highlighting significant variations among solvents and different suppliers of hBN.
The highest yield for SPEs, although still small in absolute quantities, has been observed for hBN nanoflakes from Merck (Sigma-Aldrich) while drop-casted in acetone solution.
These specific SPEs exhibit remarkable brightness ($>300$\,kCounts/s) alongside high purity ($g^{(2)}<0.15$), which are comparable to SPEs derived from exfoliated flakes \cite{Grosso2017,Gan2022}.\\
This study thus offers a systematic process analysis for the practical incorporation of SPEs into quantum applications via drop-casting. 
It demonstrates the feasibility of identifying and enhancing bright and pure quantum sources by employing appropriately chosen solvents.

\section{Results and Discussion}
\noindent In this study we investigate the generation of SPEs through drop-casting, examining the influence of both the immersion solution and the supplier of hBN.
We provide a systematic workflow analysis of drop-casted hBN nanoflakes, starting with an evaluation of their spatial distribution on the substrate before conducting photoluminescence measurements using a wide-field microscope.
These preliminary tests enable an initial estimation of potential SPEs without the need for time-consuming analysis of individual nanoflakes.
We then perform a comprehensive characterization of the optical properties of ensembles of individual emitters using a confocal microscope, enabling a thorough assessment of the measured SPE' quality.
Furthermore, specific yields are provided for distinct combinations of the hBN supplier and the immersion solution. \\   
We have developed a simple recipe that reliably generates SPEs from immersed hBN nanoflakes, which is written in detail under Section \ref{sec:meth}.
In summary, nanoflakes of hBN are immersed in solution, sonicated for about 20\,min and then drop-casted onto a silicon dioxide (SiO$_2$) wafer.
The droplet is dried up to 14\,hs overnight under ambient conditions and subsequently annealed at 870$^{\circ}$C for 15\,mins under Argon environment using a rapid thermal annealer (Fig.~1).

\begin{figure*}[h!] % [b!] for bottom
    \includegraphics[width=\textwidth]{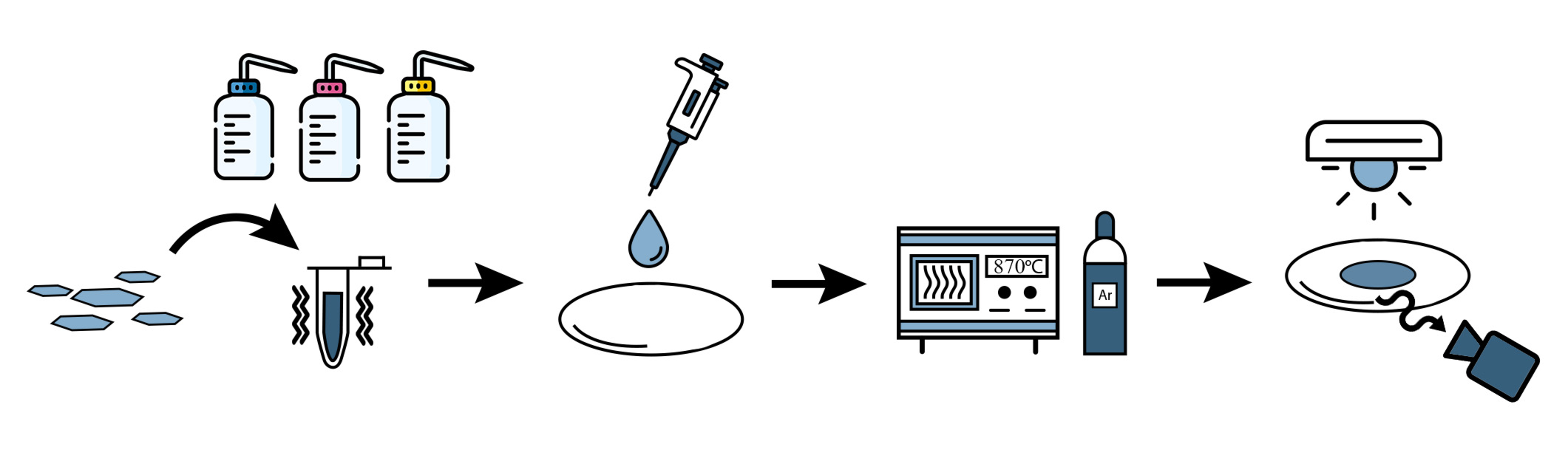}
    \caption{
    \footnotesize
    {\bf Drop-casting procedure for hBN nanoflakes.} 
    Hexagonal boron nitride (hBN) nanoflakes are first suspended in solution and shaken in an ultrasonic bath. A small droplet ($V=1\,\mu$l) is then dropcasted onto a substrate, and dried up to 14\,hs overnight. The dried nanoflakes are then annealed under argon environment in a rapid thermal annealer at 870$^{\circ}$C. The sample is then optically characterized using spectroscopy, brightfield and time-resolved fluorescence microscopy.
    }
    \label{fig1}
\end{figure*}

\subsection{Nanoflake distribution}
\noindent The spatial distribution of deposited nanoflakes on the substrate strongly depends on the drying process of the droplet.
As larger clusters of nanoflakes may form during this process, the chances of finding those composed of only few layers, which may potentially carry a single SPE, can be drastically reduced.
To mitigate large clustering, we started by investigating the influence of the solvent on the distribution of nanoflakes.\\
During the drying of the droplet, capillary flows transport the suspended nanoflakes from the center towards the pinning line at the edge of the droplet.
While some nanoflakes are deposited within the center of the droplet (Figs.~2a,e,i), the majority tends to accumulate near its edge (Figs.~2b,f,j).
This process results in a ring-shaped distribution of nanoflakes on the substrate, known as the coffee-ring effect \cite{deegan1997}.
Using a commercial brightfield microscope (Zeiss Axio Imager M2m) we can observe how the solution composing the droplet changes the spatial distribution of nanoflakes after drying.
Our observations reveal distinct behaviors based on the immersion solvent.
Specifically, when immersed in acetone (similarly for ethanol and isopropanol, detailed in SI Fig.~S1), nanoflakes within the central region of the dried droplet demonstrate uniform distribution and remain small (Fig.~2a), which is desirable.
However, the edge possesses a rather less defined structure due to a higher accumulation of clusters (Fig.~2b).
In stark contrast, the immersion of nanoflakes in water leads to the formation of larger clusters at the center (Fig.~2e), accompanied by a pronounced manifestation of the coffee-ring effect near the droplet's edge (Fig.~2f).
This result is unsurprising since hBN sheets are hydrophobic, leading to the stacking of individual layers due to van der Waals forces.
Similar clustering has been observed in the case of methanol (SI Fig.~S1e,f).
To prevent stacking, an ionic surfactant such as sodium cholate hydrate (SC) can be added to water ($c=47$\,mg/ml), where the electrostatic repulsion between surfactant molecules hinders the attachment of single hBN layers \cite{smith2021}.
Consequently, using water and surfactant, smaller and more evenly distributed nanoflakes are found near the droplet's center (Fig.~2i), while a ring-shaped accumulation remains at the edge (Fig.~2j).
The choice between hBN suppliers, whether it is Merck's hBN powder (Fig.~2 and SI Fig.~S1) or hBN nanoflake solution from Graphene Supermarket (SI Fig.~S2), both of similar size ($d<200\,$nm), does not produce visibly different spatial distributions.\\
For a comprehensive assessment of the impact of solvent and hBN supplier material on the size distribution within a droplet, we counted the number of nanoflakes and measured their corresponding sizes (Figs.~2m,n).
Among the various solvents, water with surfactant and acetone exhibited the largest number of nanoflakes per area, followed by ethanol, and water for both hBN suppliers (from Merck (Fig.~2m) and Graphene Supermarket (Fig.~2n)).
While for Merck's hBN the size distribution peaks below 1.5\,$\mu$m for all solvents with an average width of $\sim2.2$\,$\mu$m, hBN nanoflakes from Graphene Supermarket accumulate to smaller clusters $\sim0.8$\,$\mu$m with notably narrower distributions of only $\sim0.7$\,$\mu$m.
Only in the case of water, the size distribution is biased towards larger values due to the accumulation of nanoflakes into large clusters (blue curves Figs.~2m,n).
The size distribution of isopropanol and methanol are presented in SI Fig.~S3.
Notably, acetone and water with surfactant exhibited a large number of flakes with sizes $<1\,\mu$m, marking them as highly promising candidates for potential SPE sources.\\
We then use a wide-field fluorescence microscope, which allows us to quickly identify photoluminescent flakes under a broad excitation area (see methods section~\ref{sec:meth}).
We directly compare the brightfield images (Figs.~2c,g,k and SI Figs.~S1c,g,k) of nanoflakes near the center of the dried droplet with their corresponding photoluminescence (Figs.~2d,h,l and SI Figs.~S1d,h,l) and quantified the number of emitters.
In case of drop-casting in acetone (Figs.~2c,d), many photoluminescent nanoflakes, both in clustered formation and individually (see inset), have been observed.
While for water only large clusters exhibited emission (Figs.~2g,h), the addition of surfactant revealed a large number of smaller flakes under excitation (Figs.~2k,l).\\
By comparing the amount of nanoflakes in brightfield images (Figs.~2c,g,k) with those emitting photoluminescence (Figs.~2d,h,l) we find the highest percentage of photoluminescent flakes for water and surfactant (26\%), followed by ethanol (13\%) and acetone (9\%), with only 4\% for water (percentages of other solvents can be found in SI Fig.~S4).
Although these numbers cannot differentiate SPEs and broadband emitters, they are good indicators when evaluating drop-casting methods for SPE generation.
These assessments consider not only spatial distribution but also the quantity of photoluminescent nanoflakes.
Utilizing these rapid identification methodologies, we determined water as an unsuitable candidate due to the formation of substantial clusters and a low count of photoluminescent emitting nanoflakes. In contrast, acetone, water with surfactant, and ethanol exhibited potential for generating a larger number of potential SPE.

\begin{figure*}
    \includegraphics[width=\textwidth]{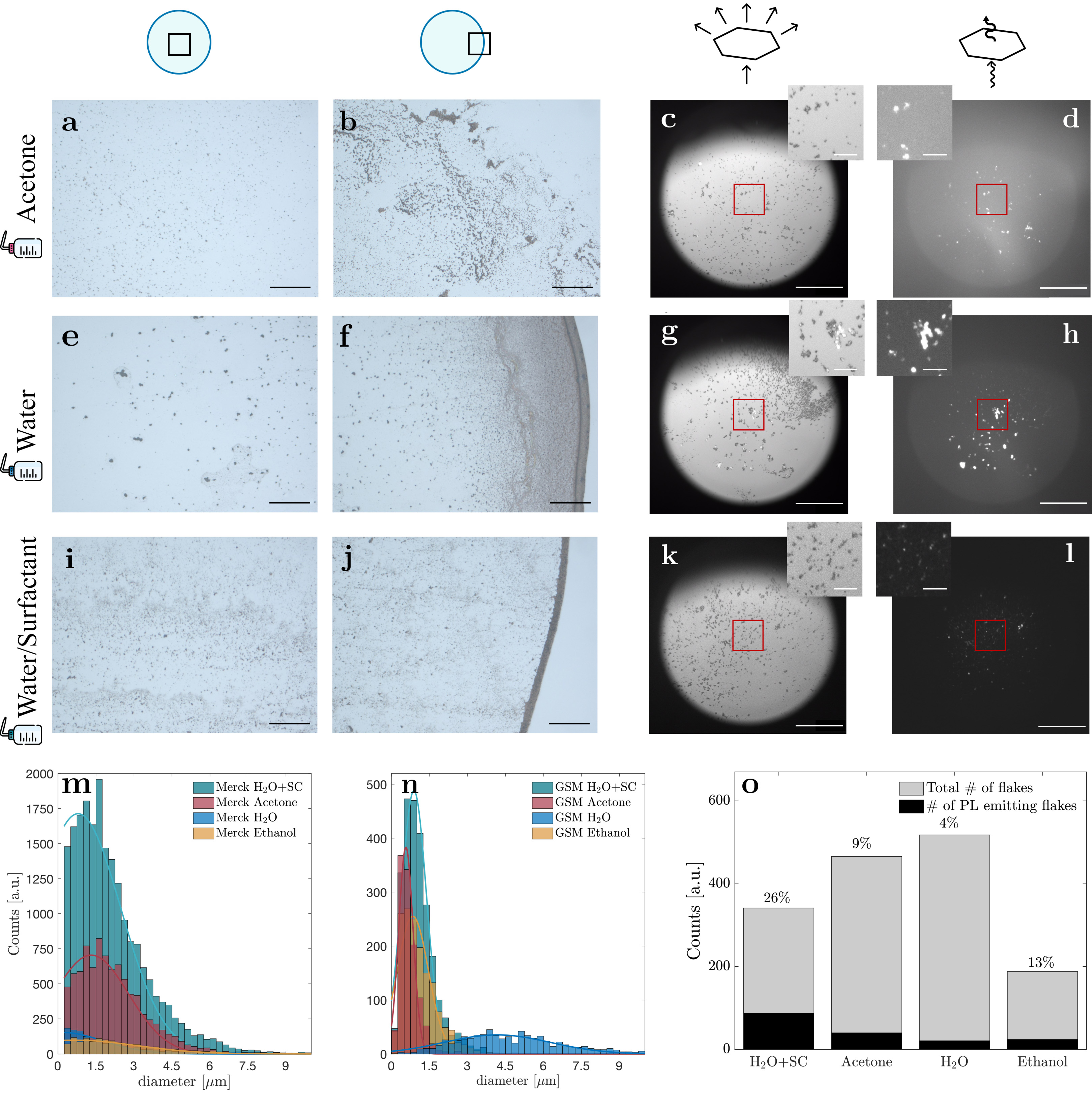}
    \caption{
    \footnotesize
    {\bf Solvent-dependence of size distribution and photoluminescence.} 
    {\bf a-d} Bright-field (BF) and photoluminescence (PL) images with acetone as solvent, {\bf e-h} for water, {\bf i-l} and for water with surfactant.
    {\bf a,e,i} BF images of the center of the dried droplet show a more uniform distribution of deposited nanoflakes, {\bf b,f,j} compared to its edge, where clusters strongly accumulate. Scale bars 100\,$\mu$m. {\bf c,g,k} Direct comparison of BF images of nanoflakes, {\bf d,h,l} with their PL show large differences between solvents. Scale bars 50\,$\mu$m, inset scale bars 20\,$\mu$m. {\bf m} Size distributions of Merck's hBN nanoflakes, {\bf n} as well as for hBN from Graphene Supermarket (GSM), vary for different solvents with water (dark blue curve) being the broadest. {\bf o} Ratio of PL emitting flakes (counted in d,h,l and SI Fig.~S1d,h,l) over their total amount (counted in c,g,k and SI Fig.~S1c,g,k) for different solvents where the highest percentage was found for water and surfactant (H$_2$O+SC).
    }
    \label{fig2}
\end{figure*}

\subsection{Optical characterization}
\noindent We perform the optical characterization of hBN nanoflakes using a commercial confocal fluorescence microscope (PicoQuant MicroTime 200) equipped with a pair of single-photon avalanche detectors (SPAD) to measure emitter lifetime and correlations (further details in methods section \ref{sec:meth}).
Employing a $\lambda_{\rm exc}=530\,$nm laser with 10\,$\mu$W power for emitter excitation, we scan the substrate using a high NA objective ($\rm NA=0.9$), through which we can localize individual hBN nanoflakes ($\pm 5\,$nm)  within a $80\times80\,\mu$m area.
In Fig.~3 we show the results of Merck's hBN initially immersed in acetone.
For this characterization, we selected single nanoflakes with a transform-limited excitation spot ($\sim300$\,nm in size) and whose life time was well below 10\,ns (blue color in Fig.~3a).
Through life-time analysis, we find an average lifetime of around 4.26\,ns for this type of emitter (Fig.~3b, see details under methods \ref{sec:meth}), consistent with previous literature reports \cite{aharonovich2016,kumar2023}. 
Over an extended acquisition time of 2\,min, a stable emission rate of around 130$\pm 10$\,kHz with no signs of blinking nor bleaching, was observed (Fig.~3c).
Spectral analysis revealed a zero phonon line at around $\lambda_{\rm ZPL}=568$\,nm, accompanied by a phonon side band around $\lambda_{\rm PSB}=615$\,nm (Fig.~3d).
Emission at this wavelength suggests a defect source due to carbon atoms according to literature \cite{mendelson2021,Kumar2024}.
However, slight peak positions variations between measured nanoflakes (Fig.~4c) indicate the potential occurrence of multiple defect types.
To verify the nanoflakes as SPEs, we have performed second-order correlation measurements $g^{(2)}(\tau)$, revealing low correlations of $g^{(2)}(\tau=0)=0.11$, indicative of single photon sources (Fig.~3e).
Further investigations involved examining the polarization dependence of excitation and emission of the SPEs by placing polarizers in the respective beam paths (see details in Section \ref{sec:meth}).
By rotating the excitation polarization of the laser beam we find typical dipole emissions at 90, 270$^{\circ}$ (Fig.~3f) by fitting the intensities using Eqn.\,\ref{EqnPol} \cite{Kumar2024}.
%We fit the intensities using Eqn.\,\ref{EqnPol}, where we find a negligible tilt of the dipole axis of less than 2$^{\circ}$.
We find similar results for the polarization dependence of the single photon emission (Fig.~3g), and whose axis aligns with that of the excitation.

\begin{figure*}[b!]
    \includegraphics[width=\textwidth]{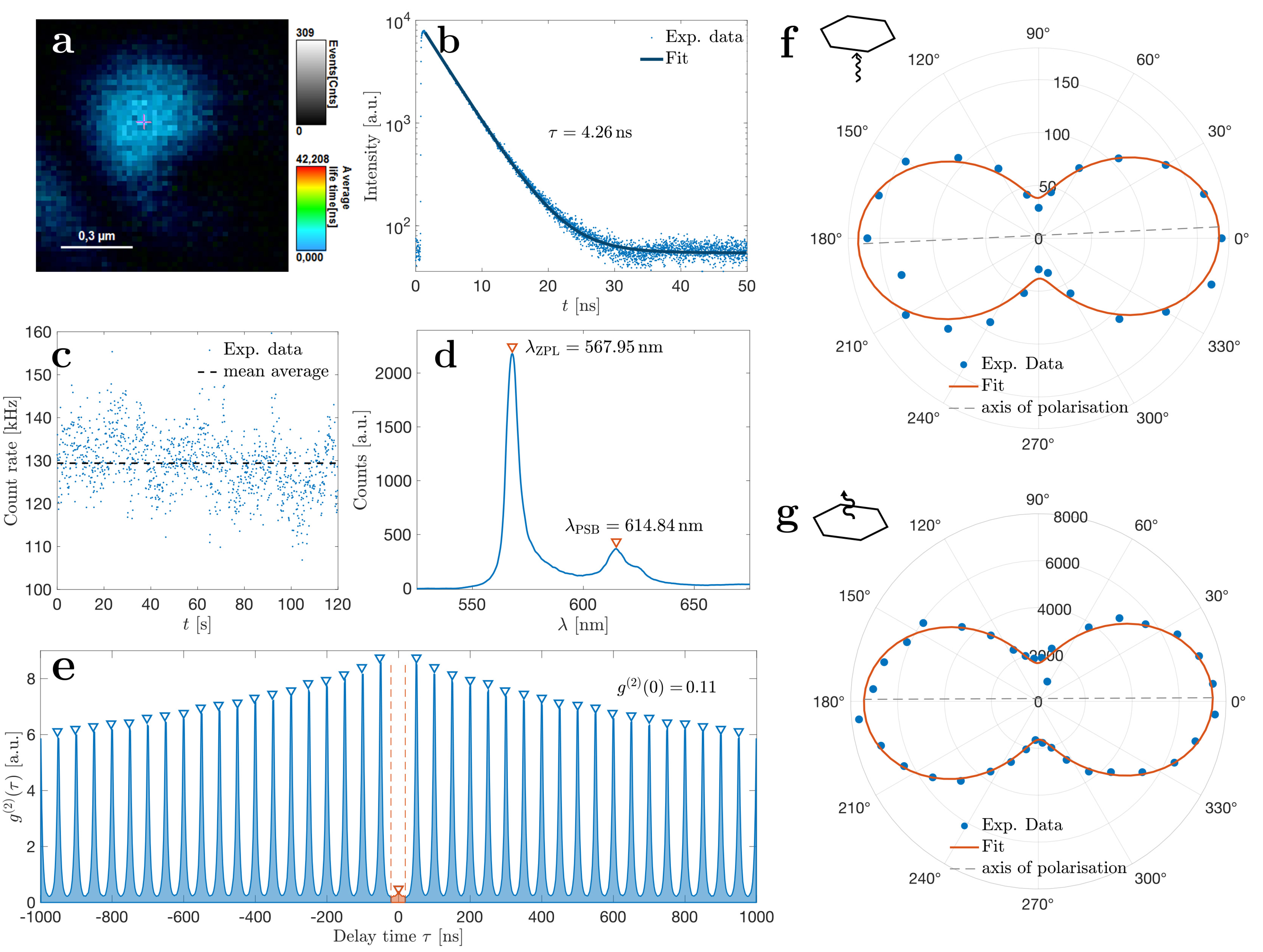}
    \caption{
    \footnotesize
    {\bf Optical characterization of Merck's hBN dropcasted in acetone.} 
    {\bf a} Scanned PL image of a nanoflake shows high brightness and short life time (blue color). {\bf b} Life-time analysis of the emitter reveal a decay time of $\tau=4.26$\,ns using Eqn.~\ref{eq:lt}. {\bf c} The emitters show constant high brightness at $130\pm8$\,kHz over 120\,s. {\bf d} Spectrum analysis shows a narrow zero phonon line at $\lambda_{\rm ZPL}=568$\,nm with a smaller phonon side band at $\lambda_{\rm PSB}=615$\,nm. {\bf e} Second-order time-correlated measurements $g^{(2)}$ show a strong dip at $\tau=0$ with $g^{(2)}=0.11$, identifying these nanoflakes as SPE. {\bf f} Excitation polarization measurements on the PL signal, {\bf g} as well as emission polarisation measurements, reveal the dipole character of the flakes.
    Both, excitation and emission polarization axis, align very well with each other.}
    \label{fig3}
\end{figure*}

\subsection{Occurrence and quality of SPE}
Following the detailed optical characterization of a single type of emitter, i.e. Merck's hBN in acetone, we have extended our analysis to assess ensembles of emitters resulting from various combinations of solvent with hBN sources.
By measuring the correlation $g^{(2)}$ at time delay zero, we discriminate between multi-photon emitters ($g^{(2)}>0.5$) and single-photon emitters ($g^{(2)}\le0.5$), thus determining the proportion of SPEs among all measured photoluminescent nanoflakes after drop-casting.
For Merck's hBN, acetone demonstrated the highest yield with 18 SPEs out of 132 measured nanoflakes (14\%), followed by water with surfactant with 10/132 (8\%), and only 7/166 (4\%) for ethanol (Fig.~4a).
Conversely, with hBN nanoflakes from Graphene Supermarket, slightly lower numbers were observed with only 3/30 (10\%) for ethanol, 9/117 (8\%) for acetone, while none were identified as SPEs for water with surfactant (Fig.~4a).
These results underscore the expected variations in SPE yield depending on the solvent used for drop-casting \cite{Ronceray2023}.
However, our results shows that drop-casting in acetone provides better SPE yields compared to ethanol, which has typically been identified as a good candidate \cite{Neumann2023,islam2023} and, in fact, is often used as medium for the shipment of hBN nanoflakes. We attribute this variation to a combination of effects; first, a difference in physical characteristics such as surface tension lead to varying distributions of nanoflakes on the substrate \cite{marsh2015,martinez2023}, and second, the activation of defect centers is affected by the specific type of organic molecule \cite{Ronceray2023}. The latter effect has not been studied yet after the drying of the liquid and subsequent thermal annealing, which shows that further studies on defect generation using different solvents are required. \\
To better understand the statistical variations within the same solvent and supplier, we plotted the measured intensities against $g^{(2)}$ values for all combinations (Fig.~4b).
By delineating the space into quadrants, the upper left quadrant, characterized by high intensities and low correlation values, is deemed most suitable for quantum applications.
Notably, the combination of acetone with Merck's hBN emerged as the sole option fulfilling both criteria, positioning it as the most suitable candidate for drop-casting SPEs (details on SPE yields and quality assessments for all measured solutions are available in SI Fig.~S5).
Furthermore, we record multiple spectra within each solvent and supplier to identify peak emissions and determine the zero phonon line.
Emissions spanning $\lambda_{\rm{em}}=550-610$\,nm (Fig.~4c) have been identified with the presence of multiple types of defect centers.
The highest occurrence of SPE emission is found around 570\,nm indicating a defect center likely due to carbon impurities in the lattice structure of hBN \cite{mendelson2021,Kumar2024}.
Although oxygen defects in hBN could produce central phonon line emission at similar wavelengths \cite{li2022}, it is less likely to occur given that the annealing occurred in the absence of oxygen.
Interestingly, minimal variation in this range was observed among acetone, ethanol, isopropanol, and methanol or between different suppliers (SI Figs.~S6), implying insignificant alteration in defect centers across these variations.
Finally, we show the actual yield of SPEs among deposited nanoflakes on the substrate (Fig.~4d).
This metric combines the count of PL emitting flakes among all deposited flakes (Fig.~2o) with the percentage of SPEs (Fig.~4a).
We find the highest number of SPEs when dropcasted in water and surfactant (H$_2$O+SC) with about 2\%.
However, when distinguishing between different qualities of SPEs according to the quadrants in Fig.~4b, only acetone demonstrated bright SPEs with a low $g^{(2)}$ value of about 0.25\%.

\begin{figure*}[b!]
    \includegraphics[width=.7\textwidth]{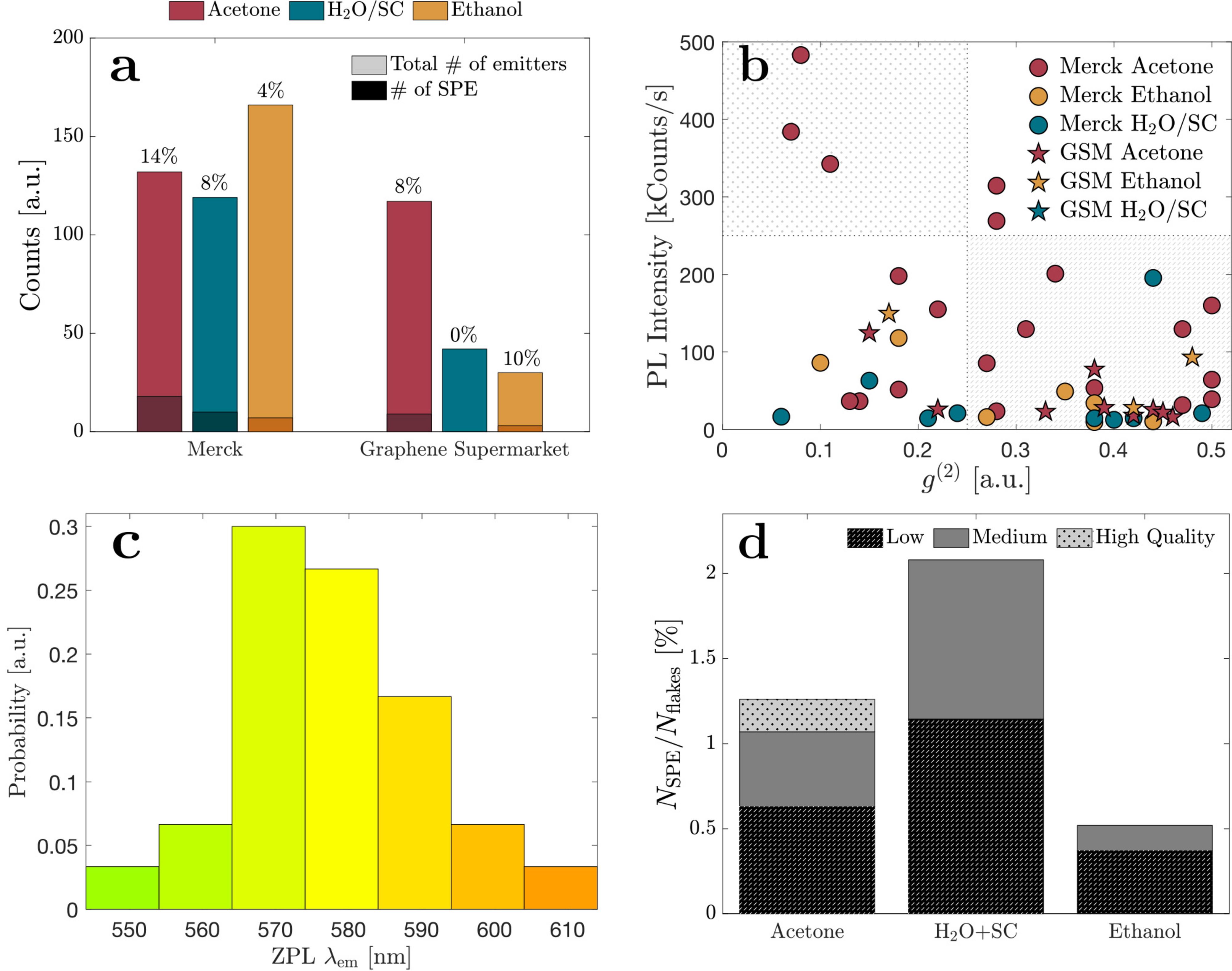}
    \caption{
    \footnotesize
    {\bf Occurence and quality of SPE.} 
    {\bf a} Ratio of SPEs to the total number of measured emitters show that the solvent acetone produces the highest number of SPEs in Merck, while ethanol is best for Graphene Supermarket. {\bf b} Map of PL intensity over $g^{(2)}$ spaced in four quadrants show that acetone and Merck's hBN combined produce the brightest and lowest $g^{(2)}$ SPE. {\bf c} SPE spectra for all combinations of solvent and hBN supplier reveal different zero phonon line (ZPL) emission between 550\,nm and 640\,nm with the highest occurrence around 570\,nm. {\bf d} Summarizes the amount of SPEs identified among all deposited flakes and  categorized depending on their quality. Highest quality (dotted quadrant in {\bf b}) for emitters that are bright and posses low $g^{(2)}$, medium for low intensity/low $g^{(2)}$ or high intensity/high $g^{(2)}$ (blank quadrants in {\bf b}), and low quality for low intensity and high $g^{(2)}$ (hatched quadrant in {\bf b}).
    }
    \label{fig4}
\end{figure*}

\subsection{Discussion on choice of solvent and hBN supplier}
\noindent The initial size of the suspended hBN nanoflakes is a crucial starting point.
For our study, we deliberately selected nanoscopic flakes that are likely comprised of a single or few layers only. 
While larger hBN flakes exceeding 1\,$\mu$m (such as 2D Semiconductor's hBN powder) could theoretically be utilized, extensive ultrasonication at high powers (250W) and prolonged suspension in water and surfactant failed to disperse these larger flakes and clusters efficiently (see SI Fig.~S7).
Consequently, only a negligible amount of nanoflakes were present ($d<1\,\mu$m), challenging their identification due to substantially weaker photoluminescence signals compared to adjacent larger flakes and clusters.
Consequently, hBN nanoflakes, whether pre-suspended in a solvent or in powder form, emerge as more suitable for drop-casting deposition.\\
Additionally, the choice of solvent significantly impacts the individual nanoflake deposition on the substrate.
Varying surface tension of the droplet and solvent interaction with the hydrophobic surfaces of hBN nanoflakes can mitigate cluster formation, ensuring a more uniform deposition, particularly in the central area of the dried droplet \cite{marsh2015}.
Optimal results were attained with water and surfactant, acetone, and ethanol, whereas pure water exhibited improper solvent behavior due to substantial clustering of flakes.
Techniques like altering capillary flows or unpinning the droplet's contact line could further alleviate the coffee-ring effect \cite{kumar2020}.
Furthermore, the photoluminescence signal from hBN flakes significantly relies on their deposition, resulting in stronger signals from nanoflakes (e.g., water and surfactant, acetone, and ethanol) compared to larger clusters (e.g., water alone).\\
While size distribution and photoluminescence emission serve as good indicators, they are not comprehensive criteria for evaluating SPE creation.
A full optical characterization of PL emitting nanoflakes was required to further distinguish SPEs from multiphoton emitters.
The choice of solvent and hBN supplier significantly influenced SPE yield and quality, notably in intensity and purity. 
Among all tested solvents and hBN suppliers, only Merck's hBN in acetone yielded significant quantity and quality of SPEs.
It is noteworthy that we exclusively considered photostable SPEs, exhibiting no bleaching or blinking during the entire measurement period (approximately 2\,min).\\
%Annealing notably enhanced emitter stability (see Methods section \ref{sec:meth}).\\
%\textcolor{red}{Discussion on what determines the intensity and g2 values of SPE}
However, only by combining both criteria, PL emission of deposited nanoflakes and SPE generation, can we accurately determine a realistic yield of SPEs through drop-casting.
While our identified numbers remain low, they align well with qualitative observations from prior drop-casting experiments \cite{chen2021Solvent,schauffert2022}.
Our stepwise analysis underscores that these low figures are not unexpected, given cluster formation during drop-casting (see Fig.~2), while not all nanoflakes will carry defect centers that emit under $\lambda_{\rm exc}=530\,$nm excitation.
Moreover, as the emitters highly depend on the excitation polarization of the laser beam (Fig.~3f), not all SPEs will be visible under fixed linear polarization, potentially increasing the actual count.
The fundamental findings of this work, however, remain unaffected by these experimental details.

\section{Conclusions and Outlook}
\label{sec:Concl}
\noindent We have shown a systematic approach to study the generation of SPEs from drop-casted hBN nanoflakes demonstrating the feasibility of the methods and providing a realistic yield assessment.
Our results allow for a better evaluation of current drop-casting methods compared to deterministic exfoliation and transfer methods.\\
The choice of solvent, alongside the supplied hBN material, notably influences SPE yield and quality.
This holds particular significance for employing drop-casted nanoflakes as SPE sources in quantum communication, sensing, and imaging.
The versatility of the drop-casting approach and significant reduction of involved equipment and manual labour allow for a much faster and seamless integration of SPEs into nanophotonic systems.
For example, hBN drop-casting facilitates flake deposition on the tip of a single-mode fiber (SI Figs.~S8a-c) or on more complex structures like those of exposed core fibers (SI Fig.~8d), which are difficult to access with conventional transfer techniques.\\
However, hurdles remain for drop-casting to become a practical solution for creating SPE sources, as the elimination of undesired emitters is essential to ensure only intended SPEs populate a given nanophotonic system.
This necessitates additional steps like laser ablation \cite{ngo2022} or secondary transfer techniques.
Nevertheless, drop-casting offers an alternative, highly controllable means of depositing hBN flakes.

\clearpage
\section{Methods}
\label{sec:meth}

\subsection{Optical setups}
\noindent \textbf{Time-resolved confocal microscope}\\
The optical investigation of quantum emitters is carried out using a commercial time-resolved confocal microscope (PicoQuant MicroTime 200).
An excitation laser with wavelength 530\,nm, pulse rate 20\,MHz and excitation power of around 10\,$\mu$W (measured before the objective) was used.
The detection path of the setup uses various filters (long- and bandpass) to block the excitation laser.
The setup is equipped with two SPADs (single-photon avalanche diodes), producing a Hanbury-Brown-Twist (HBT) interferometer.
The photoluminescence map of each nanoflake is created by scanning the sample stage with 2\,ms dwell time and a laser repetition rate of 20\,MHz.
The photoluminescence signal is collected using a 100$\times$ dry immersion objective with a high numerical aperture (NA) of 0.9.
Through the 50:50 beamsplitter, the emission signal is split into two arms.
At the end of each path, a SPAD is placed to measure the second-order correlation function.
The data analysis of the correlation function, as well as lifetime measurements, is performed with built-in software.
The sample is rastered in $x-y$ using a piezoscanner attached to a sample holder with nanometer precision.
Depending on the acquisition mode (fast/slow) the spatial resolution can be increased from 20\,nm down to 5\,nm.
The data acquisition time for these measurements was around 2\,min per emitter.
A spectrometer (Andor Kymera 328i-D2-sil) is attached to one of the exit ports of the optical setup to collect the emitter's spectra.\\

\noindent \textbf{Photoluminescence wide-field microscope}\\
To measure the photoluminescence of a large area, a custom-built fluorescence microscope was used (SI Fig.~S9).
A light beam from a 530\,nm laser is focused with a field lens onto the back focal plane of a 50$\times$ long working distance microscope objective (Zeiss EC Epiplan Neofluar 50$\times$/0.55).
This results in the distribution of the laser radiation on a larger area compared to the focus of a collimated beam.
A longpass dichroic mirror (Thorlabs DMLP550R) separates the signal from the excitation laser radiation, with additional filtering provided by a long-pass filter (2x Thorlabs FELH0550).
The microscope images are formed onto a back-illuminated scientific CMOS camera (Teledyne Photometrics Kinetix) using a tube lens.
To view the sample, a white-light illumination is also available (not shown in SI Fig.~S9).

% Large clusters, strongly present in the case of water (Figs.~2g,h), are found to be the brightest, while the sensitivity of this setup allows for identification of photoluminescent nanoflakes above $d=500$\,nm.

\subsection{Sample preparation}
\noindent We have obtained hBN nanoflakes as powder from Merck (790532, SigmaAldrich) with an average particle size $<150\,$nm, and as water/ethanol solution from Graphene Supermarket with lateral sizes between 50-200\,nm.
In case of hBN powder, a small quantity $<1\,$mg was immersed in 2\,ml solution and subsequently shaken in an ultrasonic bath (100\,W) between 20-30\,mins.
For hBN nanoflakes already in solution, an additional drying step was added before reimmersion in solution. 
We have investigated the influence of six solutions on the drop-casting of hBN: acetone ($>99\%$ purity), ethanol ($>99\%$ purity), isopropanol ($>70\%$ purity), DI water, and a mixture of H$_2$O with sodium cholate hydrate (SC, $c=47$\,mg/ml) acting as efficient surfactant.
A drop of 1$\mu$l of solution is then dropcasted onto a SiO$_2$ wafer, and dried under ambient conditions overnight. This ensures that all remaining solution has been evaporated before annealing.
The sample is then annealed following standard procedures (REF) at 870$^{\circ}$C for 15\,mins under Argon atmosphere in a rapid thermal annealer (RTA, Allwin21 Corp. AccuTherm AW610M). Before annealing, we flush the RTA for 300\,s with nitrogen gas to avoid the creation of new defects in hBN in an oxygen environment.

\subsection{Particle counting}
\noindent For the determination of the size distribution, the drop-casted nanoflakes have been measured via particle counting.
For this, the background has been subtracted from the image and thresholding applied. 
The amount of flakes that emit under photoluminescence, are counted after thresholding for flakes with an area 0.15-3.7\,$\mu$m$^2$.
This ensures that no noise from the background or large clusters are counted.

\subsection{Life-time analysis}
\noindent The lifetime of the emitters $t$ is extracted from the pulsed $g^{(2)}$ measurement by fitting the experimental data with an exponential decay function

\begin{equation}
f(x) = C + \sum^n_{i=-n} A_i\times e^{\frac{\tau-i\cdot p}{t}},
\label{eq:lt}
\end{equation}
where $C$ is a constant offset in the presence of noise, $A_{i,j}$  the coefficients for growth and decay respectively (see also SI Fig.~S10).
Each peak is fitted due to the finite measurement time and re-emission peak dynamics.

\subsection{Polarisation dependent measurements}
\noindent For the polarization-resolved measurements, we extended the capabilities of the commercial fluorescence lifetime microscope (PicoQuant MicroTime 200) by inserting polarising elements in the setup, which include: (i) a fixed linear polarizer in the excitation laser path to obtain horizontal polarization with a high extinction ratio, (ii) a quarter-wave plate for circularly polarised light, (iii) a polarizer that scans the emitter excitation axis after the quarter-wave plate, and (iv) a polarizer in the detection path for measuring the polarization of emission.
All polarizing elements were controlled via a motorized rotation mount (Thorlabs ELL14).
During these polarization measurements the laser power remained constant (SI Fig.~S11).
The measured angular dependent intensity of the emitter is fitted with cosine squared-function (see Eq.\ref{EqnPol}) to extract the axis of polarization.

\begin{equation}
I(\theta) = a \cdot \cos^2(\theta - b) + c,
\end{equation}
\label{EqnPol}

\noindent where a, b and c are the fitting parameters with $\theta$ as axis of polarization. The degree of (linear) polarization is defined as 

\begin{equation}
    \text{DOP} = \frac{ \text{I}_\text{max} - \text{I}_\text{min}}{\text{I}_\text{max} + \text{I}_\text{min}}.
    \label{dopeq}
\end{equation}

\begin{acknowledgments}
\noindent We thank A. Nowotnick and C. Ronning for providing us with additional hBN material. F.E., F.S. and S.R. acknowledge support by the German Federal Ministry of Education and Research (BMBF) via the project “tubLAN Q.0” (Grant No. 13N14876). This project was supported by the European Union, the European Social Funds and the Federal State of Thuringia as FGI 0043 under grant ID 2021FGI0043. A.K. and T.V. acknowledge support by the Deutsche Forschungsgemeinschaft (DFG, German Research Foundation) - Projektnummer 445275953 and by the German Space Agency DLR with funds provided by the Federal Ministry for Economic Affairs and Climate Action BMWK under grant number 50WM2165 (QUICK3) and 50RP2200 (QuVeKS). T.V. was funded by the Federal Ministry of Education and Research (BMBF) under grant number 13N16292.

\end{acknowledgments}

%\section{Author contributions}
%\noindent All authors have discussed the results. FS, FE planned the experiments. TK and FS fabricated the experimental samples. TK, FS, AK performed the experiments. SR built the widefield PL microscope. All authors contributed to discussions and to the writing of the article.

%\section{Additional Information}
%\vskip 0.3cm 
%\noindent\textbf{Competing interests:} The authors declare no conflict of interest.
%\vskip 0.3cm 
%\noindent\textbf{Data availability:}
%Experimental data is available upon reasonable request.
%All experimental data are available via the depository figshare under\\ \url{https://doi.org/10.6084/m9.figshare.20037581.v1}.
\vskip 0.3cm 
\noindent\textbf{Correspondence:}
Correspondence and requests for materials should be addressed to FS. 

%\appendix
\bibliographystyle{naturemag}
\bibliography{ref_hBN}

\begin{thebibliography}{10}
\expandafter\ifx\csname url\endcsname\relax
  \def\url#1{\texttt{#1}}\fi
\expandafter\ifx\csname urlprefix\endcsname\relax\def\urlprefix{URL }\fi
\providecommand{\bibinfo}[2]{#2}
\providecommand{\eprint}[2][]{\url{#2}}

\bibitem{aharonovich2016}
\bibinfo{author}{Aharonovich, I.}, \bibinfo{author}{Englund, D.} \&
  \bibinfo{author}{Toth, M.}
\newblock \bibinfo{title}{Solid-state single-photon emitters}.
\newblock \emph{\bibinfo{journal}{Nature photonics}}
  \textbf{\bibinfo{volume}{10}}, \bibinfo{pages}{631--641}
  (\bibinfo{year}{2016}).

\bibitem{zeng2022}
\bibinfo{author}{Zeng, H. Z.~J.} \emph{et~al.}
\newblock \bibinfo{title}{Integrated room temperature single-photon source for
  quantum key distribution}.
\newblock \emph{\bibinfo{journal}{Optics Letters}}
  \textbf{\bibinfo{volume}{47}}, \bibinfo{pages}{1673--1676}
  (\bibinfo{year}{2022}).

\bibitem{steinlechner2013}
\bibinfo{author}{Steinlechner, S.} \emph{et~al.}
\newblock \bibinfo{title}{Quantum-dense metrology}.
\newblock \emph{\bibinfo{journal}{Nature Photonics}}
  \textbf{\bibinfo{volume}{7}}, \bibinfo{pages}{626--630}
  (\bibinfo{year}{2013}).

\bibitem{kok2007}
\bibinfo{author}{Kok, P.} \emph{et~al.}
\newblock \bibinfo{title}{Linear optical quantum computing with photonic
  qubits}.
\newblock \emph{\bibinfo{journal}{Reviews of modern physics}}
  \textbf{\bibinfo{volume}{79}}, \bibinfo{pages}{135} (\bibinfo{year}{2007}).

\bibitem{zerom2011}
\bibinfo{author}{Zerom, P.}, \bibinfo{author}{Chan, K. W.~C.},
  \bibinfo{author}{Howell, J.~C.} \& \bibinfo{author}{Boyd, R.~W.}
\newblock \bibinfo{title}{Entangled-photon compressive ghost imaging}.
\newblock \emph{\bibinfo{journal}{Physical Review A}}
  \textbf{\bibinfo{volume}{84}}, \bibinfo{pages}{061804}
  (\bibinfo{year}{2011}).

\bibitem{Tran2016Q}
\bibinfo{author}{Tran, T.~T.}, \bibinfo{author}{Bray, K.},
  \bibinfo{author}{Ford, M.~J.}, \bibinfo{author}{Toth, M.} \&
  \bibinfo{author}{Aharonovich, I.}
\newblock \bibinfo{title}{Quantum emission from hexagonal boron nitride
  monolayers}.
\newblock \emph{\bibinfo{journal}{Nature Nanotechnology}}
  \textbf{\bibinfo{volume}{11}}, \bibinfo{pages}{37--41}
  (\bibinfo{year}{2016}).

\bibitem{tran2016}
\bibinfo{author}{Tran, T.~T.} \emph{et~al.}
\newblock \bibinfo{title}{Robust multicolor single photon emission from point
  defects in hexagonal boron nitride}.
\newblock \emph{\bibinfo{journal}{ACS nano}} \textbf{\bibinfo{volume}{10}},
  \bibinfo{pages}{7331--7338} (\bibinfo{year}{2016}).

\bibitem{vogl2017}
\bibinfo{author}{Vogl, T.}, \bibinfo{author}{Lu, Y.} \& \bibinfo{author}{Lam,
  P.~K.}
\newblock \bibinfo{title}{Room temperature single photon source using
  fiber-integrated hexagonal boron nitride}.
\newblock \emph{\bibinfo{journal}{Journal of Physics D: Applied Physics}}
  \textbf{\bibinfo{volume}{50}}, \bibinfo{pages}{295101}
  (\bibinfo{year}{2017}).

\bibitem{sajid2020}
\bibinfo{author}{Sajid, A.}, \bibinfo{author}{Ford, M.~J.} \&
  \bibinfo{author}{Reimers, J.~R.}
\newblock \bibinfo{title}{Single-photon emitters in hexagonal boron nitride: a
  review of progress}.
\newblock \emph{\bibinfo{journal}{Reports on Progress in Physics}}
  \textbf{\bibinfo{volume}{83}}, \bibinfo{pages}{044501}
  (\bibinfo{year}{2020}).

\bibitem{vogl2018}
\bibinfo{author}{Vogl, T.}, \bibinfo{author}{Campbell, G.},
  \bibinfo{author}{Buchler, B.~C.}, \bibinfo{author}{Lu, Y.} \&
  \bibinfo{author}{Lam, P.~K.}
\newblock \bibinfo{title}{Fabrication and deterministic transfer of
  high-quality quantum emitters in hexagonal boron nitride}.
\newblock \emph{\bibinfo{journal}{ACS Photonics}} \textbf{\bibinfo{volume}{5}},
  \bibinfo{pages}{2305--2312} (\bibinfo{year}{2018}).

\bibitem{chen2021}
\bibinfo{author}{Chen, Y.} \emph{et~al.}
\newblock \bibinfo{title}{Generation of high-density quantum emitters in
  high-quality, exfoliated hexagonal boron nitride}.
\newblock \emph{\bibinfo{journal}{ACS Applied Materials \& Interfaces}}
  \textbf{\bibinfo{volume}{13}}, \bibinfo{pages}{47283--47292}
  (\bibinfo{year}{2021}).

\bibitem{li2021}
\bibinfo{author}{Li, C.} \emph{et~al.}
\newblock \bibinfo{title}{Scalable and deterministic fabrication of quantum
  emitter arrays from hexagonal boron nitride}.
\newblock \emph{\bibinfo{journal}{Nano Letters}} \textbf{\bibinfo{volume}{21}},
  \bibinfo{pages}{3626--3632} (\bibinfo{year}{2021}).

\bibitem{Choi2016}
\bibinfo{author}{Choi, S.} \emph{et~al.}
\newblock \bibinfo{title}{Engineering and localization of quantum emitters in
  large hexagonal boron nitride layers}.
\newblock \emph{\bibinfo{journal}{ACS Applied Materials \& Interfaces}}
  \textbf{\bibinfo{volume}{8}}, \bibinfo{pages}{29642--29648}
  (\bibinfo{year}{2016}).

\bibitem{Chejanovsky2016}
\bibinfo{author}{Chejanovsky, N.} \emph{et~al.}
\newblock \bibinfo{title}{Structural attributes and photodynamics of visible
  spectrum quantum emitters in hexagonal boron nitride}.
\newblock \emph{\bibinfo{journal}{Nano Letters}} \textbf{\bibinfo{volume}{16}},
  \bibinfo{pages}{7037--7045} (\bibinfo{year}{2016}).

\bibitem{Vogl2019}
\bibinfo{author}{Vogl, T.} \emph{et~al.}
\newblock \bibinfo{title}{Radiation tolerance of two-dimensional material-based
  devices for space applications}.
\newblock \emph{\bibinfo{journal}{Nature Communications}}
  \textbf{\bibinfo{volume}{10}}, \bibinfo{pages}{1202} (\bibinfo{year}{2019}).

\bibitem{Gao2021}
\bibinfo{author}{Gao, X.} \emph{et~al.}
\newblock \bibinfo{title}{Femtosecond laser writing of spin defects in
  hexagonal boron nitride}.
\newblock \emph{\bibinfo{journal}{ACS Photonics}} \textbf{\bibinfo{volume}{8}},
  \bibinfo{pages}{994--1000} (\bibinfo{year}{2021}).

\bibitem{kumar2023}
\bibinfo{author}{Kumar, A.} \emph{et~al.}
\newblock \bibinfo{title}{Localized creation of yellow single photon emitting
  carbon complexes in hexagonal boron nitride}.
\newblock \emph{\bibinfo{journal}{APL Materials}}
  \textbf{\bibinfo{volume}{11}}, \bibinfo{pages}{071108}
  (\bibinfo{year}{2023}).

\bibitem{hausler2021}
\bibinfo{author}{H{\"a}u{\ss}ler, S.} \emph{et~al.}
\newblock \bibinfo{title}{Tunable fiber-cavity enhanced photon emission from
  defect centers in hbn}.
\newblock \emph{\bibinfo{journal}{Advanced Optical Materials}}
  \textbf{\bibinfo{volume}{9}}, \bibinfo{pages}{2002218}
  (\bibinfo{year}{2021}).

\bibitem{Vogl:2019aa}
\bibinfo{author}{Vogl, T.}, \bibinfo{author}{Lecamwasam, R.},
  \bibinfo{author}{Buchler, B.~C.}, \bibinfo{author}{Lu, Y.} \&
  \bibinfo{author}{Lam, P.~K.}
\newblock \bibinfo{title}{Compact cavity-enhanced single-photon generation with
  hexagonal boron nitride}.
\newblock \emph{\bibinfo{journal}{ACS Photonics}} \textbf{\bibinfo{volume}{6}},
  \bibinfo{pages}{1955--1962} (\bibinfo{year}{2019}).

\bibitem{chen2021bottom}
\bibinfo{author}{Chen, Y.} \emph{et~al.}
\newblock \bibinfo{title}{Bottom-up synthesis of hexagonal boron nitride
  nanoparticles with intensity-stabilized quantum emitters}.
\newblock \emph{\bibinfo{journal}{Small}} \textbf{\bibinfo{volume}{17}},
  \bibinfo{pages}{2008062} (\bibinfo{year}{2021}).

\bibitem{chen2021Solvent}
\bibinfo{author}{Chen, Y.} \emph{et~al.}
\newblock \bibinfo{title}{Solvent-exfoliated hexagonal boron nitride nanoflakes
  for quantum emitters}.
\newblock \emph{\bibinfo{journal}{ACS Applied Nano Materials}}
  \textbf{\bibinfo{volume}{4}}, \bibinfo{pages}{10449--10457}
  (\bibinfo{year}{2021}).

\bibitem{barelli2023}
\bibinfo{author}{Barelli, M.} \emph{et~al.}
\newblock \bibinfo{title}{Single-photon emitting arrays by capillary assembly
  of colloidal semiconductor cdse/cds/sio2 nanocrystals}.
\newblock \emph{\bibinfo{journal}{ACS photonics}}
  \textbf{\bibinfo{volume}{10}}, \bibinfo{pages}{1662--1670}
  (\bibinfo{year}{2023}).

\bibitem{schauffert2022}
\bibinfo{author}{Schauffert, H.} \emph{et~al.}
\newblock \bibinfo{title}{Characteristics of quantum emitters in hexagonal
  boron nitride suitable for integration with nanophotonic platforms}.
\newblock \emph{\bibinfo{journal}{arXiv preprint arXiv:2210.11099}}
  (\bibinfo{year}{2022}).

\bibitem{zhang2023}
\bibinfo{author}{Zhang, X.}, \bibinfo{author}{Lai, J.} \&
  \bibinfo{author}{Gray, T.}
\newblock \bibinfo{title}{Recent progress in low-temperature cvd growth of 2d
  materials}.
\newblock \emph{\bibinfo{journal}{Oxford Open Materials Science}}
  \bibinfo{pages}{itad010} (\bibinfo{year}{2023}).

\bibitem{kumar2020}
\bibinfo{author}{Kumar, A. K.~S.}, \bibinfo{author}{Zhang, Y.},
  \bibinfo{author}{Li, D.} \& \bibinfo{author}{Compton, R.~G.}
\newblock \bibinfo{title}{A mini-review: How reliable is the drop casting
  technique?}
\newblock \emph{\bibinfo{journal}{Electrochemistry Communications}}
  \textbf{\bibinfo{volume}{121}}, \bibinfo{pages}{106867}
  (\bibinfo{year}{2020}).

\bibitem{marsh2015}
\bibinfo{author}{Marsh, K.}, \bibinfo{author}{Souliman, M.} \&
  \bibinfo{author}{Kaner, R.~B.}
\newblock \bibinfo{title}{Co-solvent exfoliation and suspension of hexagonal
  boron nitride}.
\newblock \emph{\bibinfo{journal}{Chemical communications}}
  \textbf{\bibinfo{volume}{51}}, \bibinfo{pages}{187--190}
  (\bibinfo{year}{2015}).

\bibitem{smith2021}
\bibinfo{author}{Smith~McWilliams, A.~D.} \emph{et~al.}
\newblock \bibinfo{title}{Understanding the exfoliation and dispersion of
  hexagonal boron nitride nanosheets by surfactants: Implications for
  antibacterial and thermally resistant coatings}.
\newblock \emph{\bibinfo{journal}{ACS Applied Nano Materials}}
  \textbf{\bibinfo{volume}{4}}, \bibinfo{pages}{142--151}
  (\bibinfo{year}{2021}).

\bibitem{martinez2023}
\bibinfo{author}{Mart{\'\i}nez-Jim{\'e}nez, C.}, \bibinfo{author}{Chow, A.},
  \bibinfo{author}{Smith~McWilliams, A.~D.} \& \bibinfo{author}{Mart{\'\i},
  A.~A.}
\newblock \bibinfo{title}{Hexagonal boron nitride exfoliation and dispersion}.
\newblock \emph{\bibinfo{journal}{Nanoscale}} \textbf{\bibinfo{volume}{15}},
  \bibinfo{pages}{16836--16873} (\bibinfo{year}{2023}).

\bibitem{Ronceray2023}
\bibinfo{author}{Ronceray, N.} \emph{et~al.}
\newblock \bibinfo{title}{Liquid-activated quantum emission from pristine
  hexagonal boron nitride for nanofluidic sensing}.
\newblock \emph{\bibinfo{journal}{Nature Materials}}
  \textbf{\bibinfo{volume}{22}}, \bibinfo{pages}{1236--1242}
  (\bibinfo{year}{2023}).

\bibitem{islam2023}
\bibinfo{author}{Islam, M.} \emph{et~al.}
\newblock \bibinfo{title}{Large-scale statistical analysis of defect emission
  in hbn: Revealing spectral families and influence of flakes morphology}.
\newblock \emph{\bibinfo{journal}{arXiv preprint arXiv:2309.15023}}
  (\bibinfo{year}{2023}).

\bibitem{Grosso2017}
\bibinfo{author}{Grosso, G.} \emph{et~al.}
\newblock \bibinfo{title}{Tunable and high-purity room temperature
  single-photon emission from atomic defects in hexagonal boron nitride}.
\newblock \emph{\bibinfo{journal}{Nature communications}}
  \textbf{\bibinfo{volume}{8}}, \bibinfo{pages}{1--8} (\bibinfo{year}{2017}).

\bibitem{Gan2022}
\bibinfo{author}{Gan, L.} \emph{et~al.}
\newblock \bibinfo{title}{Large-scale, high-yield laser fabrication of bright
  and pure single-photon emitters at room temperature in hexagonal boron
  nitride}.
\newblock \emph{\bibinfo{journal}{ACS nano}} \textbf{\bibinfo{volume}{16}},
  \bibinfo{pages}{14254--14261} (\bibinfo{year}{2022}).

\bibitem{deegan1997}
\bibinfo{author}{Deegan, R.~D.} \emph{et~al.}
\newblock \bibinfo{title}{Capillary flow as the cause of ring stains from dried
  liquid drops}.
\newblock \emph{\bibinfo{journal}{Nature}} \textbf{\bibinfo{volume}{389}},
  \bibinfo{pages}{827--829} (\bibinfo{year}{1997}).

\bibitem{mendelson2021}
\bibinfo{author}{Mendelson, N.} \emph{et~al.}
\newblock \bibinfo{title}{Identifying carbon as the source of visible
  single-photon emission from hexagonal boron nitride}.
\newblock \emph{\bibinfo{journal}{Nature materials}}
  \textbf{\bibinfo{volume}{20}}, \bibinfo{pages}{321--328}
  (\bibinfo{year}{2021}).

\bibitem{Kumar2024}
\bibinfo{author}{Kumar, A.} \emph{et~al.}
\newblock \bibinfo{title}{Polarization dynamics of solid-state quantum
  emitters}.
\newblock \emph{\bibinfo{journal}{ACS Nano}} \textbf{\bibinfo{volume}{18}},
  \bibinfo{pages}{5270--5281} (\bibinfo{year}{2024}).

\bibitem{Neumann2023}
\bibinfo{author}{Neumann, M.} \emph{et~al.}
\newblock \bibinfo{title}{Organic molecules as origin of visible-range single
  photon emission from hexagonal boron nitride and mica}.
\newblock \emph{\bibinfo{journal}{ACS Nano}} \textbf{\bibinfo{volume}{17}},
  \bibinfo{pages}{11679--11691} (\bibinfo{year}{2023}).

\bibitem{li2022}
\bibinfo{author}{Li, S.} \& \bibinfo{author}{Gali, A.}
\newblock \bibinfo{title}{Identification of an oxygen defect in hexagonal boron
  nitride}.
\newblock \emph{\bibinfo{journal}{The Journal of Physical Chemistry Letters}}
  \textbf{\bibinfo{volume}{13}}, \bibinfo{pages}{9544--9551}
  (\bibinfo{year}{2022}).

\bibitem{ngo2022}
\bibinfo{author}{Ngo, G.~Q.} \emph{et~al.}
\newblock \bibinfo{title}{In-fibre second-harmonic generation with embedded
  two-dimensional materials}.
\newblock \emph{\bibinfo{journal}{Nature Photonics}}
  \textbf{\bibinfo{volume}{16}}, \bibinfo{pages}{769--776}
  (\bibinfo{year}{2022}).

\end{thebibliography}

\newpage

\setcounter{figure}{0}
\renewcommand{\thefigure}{S\arabic{figure}}

\section{Supplementary Information}

% Fig S1 - Schematic of the various materials involved in the experiments
\begin{figure*}[h]
    \includegraphics[width=\textwidth]{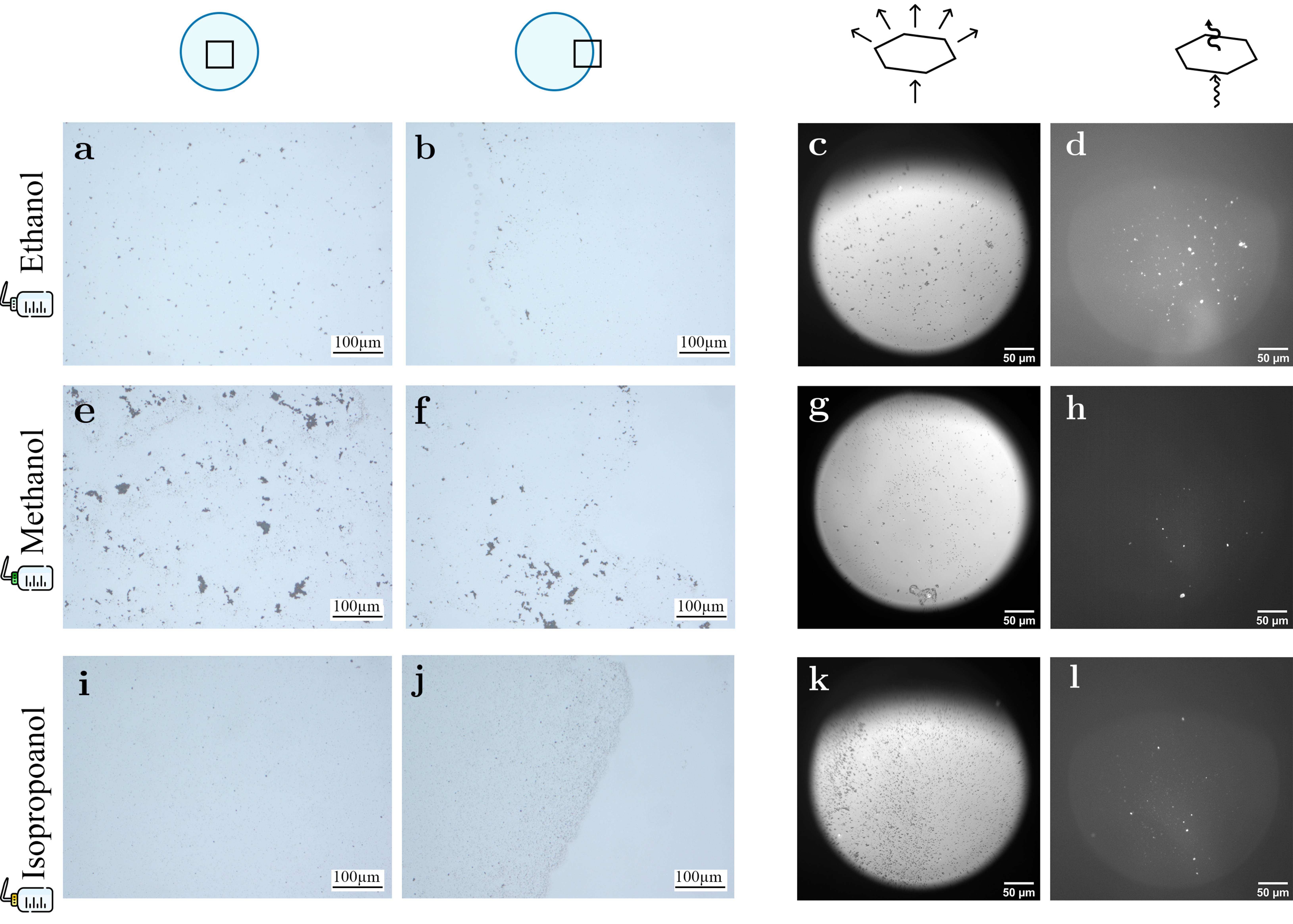}
    \caption{
    {\bf Solvent-dependence of size distribution and photoluminescence for ethanol {\bf a-d}, methanol {\bf e-h}, and isopropanol {\bf i-l} using Merck's hBN.} While for ethanol {\bf a} and isopropanol flakes are homogeneously distributed in the center of the dried droplet, for methanol {\bf e} large clusters are visible. {\bf b,f,j} Although the edge of the droplet is visible no large accumulation of flakes occur. {\bf c,g,k} Direct comparison of BF images of flakes, {\bf d,h,l} with their PL show {\bf d} more emitting flakes for ethanol, {\bf h} compared to methanol and {\bf l} isopropanol.
    }
    \label{figS1}
\end{figure*}

\clearpage

\begin{figure*}
    \includegraphics[width=\textwidth]{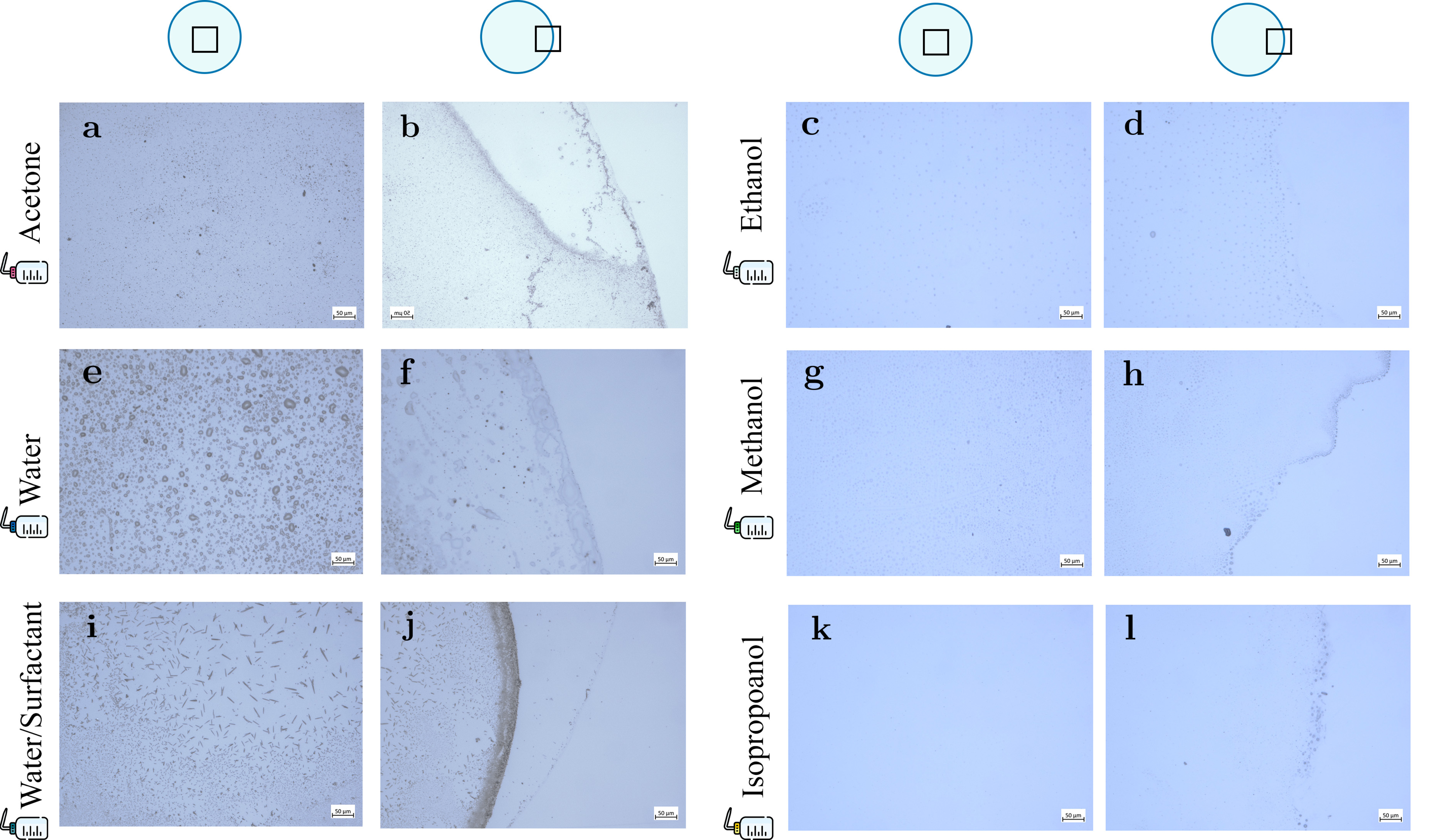}
    \caption{
    {\bf Solvent-dependence of size distribution for hBN from Graphene Supermarket.} {\bf a,b} For acetone smaller clusters are visible while they are {\bf a} homogeneously distributed across the center, {\bf b} and larger accumulations near its edge. {\bf e} For water most clusters are formed in the center, {\bf f} while few remain at its edge. {\bf i} For water and surfactant particle clusters are not observed in the center but crystals form due to residue surfactant, {\bf j} while a clear coffee-ring forms as its edge. {\bf c,d} For ethanol, {\bf g,h} methanol, and {\bf k,l} isopropanol no clusters are observed in the center while smaller accumulations are found near the edge.
    }
    \label{figS2}
\end{figure*}

\clearpage

\begin{figure*}
    \includegraphics[width=.75\textwidth]{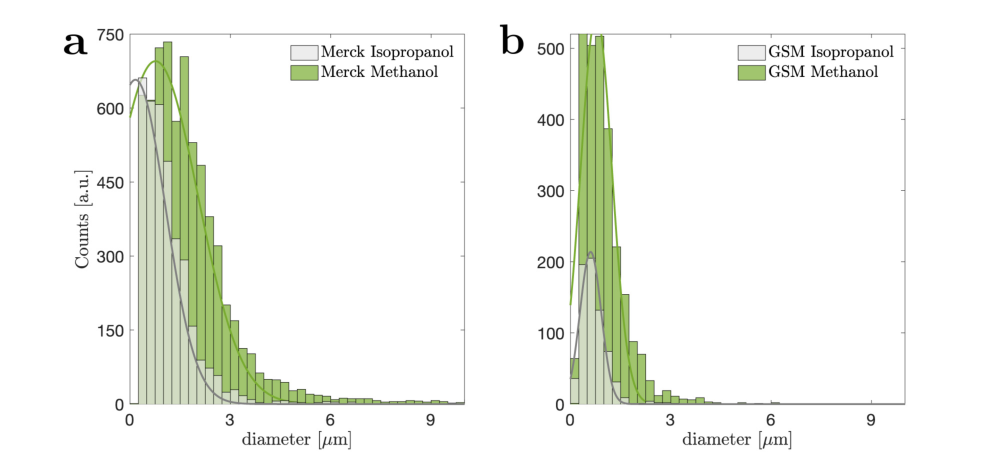}
    \caption{
    {\bf Size distribution after immersion in isopropanol and methanol.} 
    {\bf a} For hBN from Merck, the size distributions of isopropanol (white) and methanol (green), and {\bf b} for hBN from Graphene Supermarket (GSM) are all narrowly distributed below $1\,\mu$m.
    }
    \label{figS3}
\end{figure*}

\begin{figure*}
    \includegraphics[width=.5\textwidth]{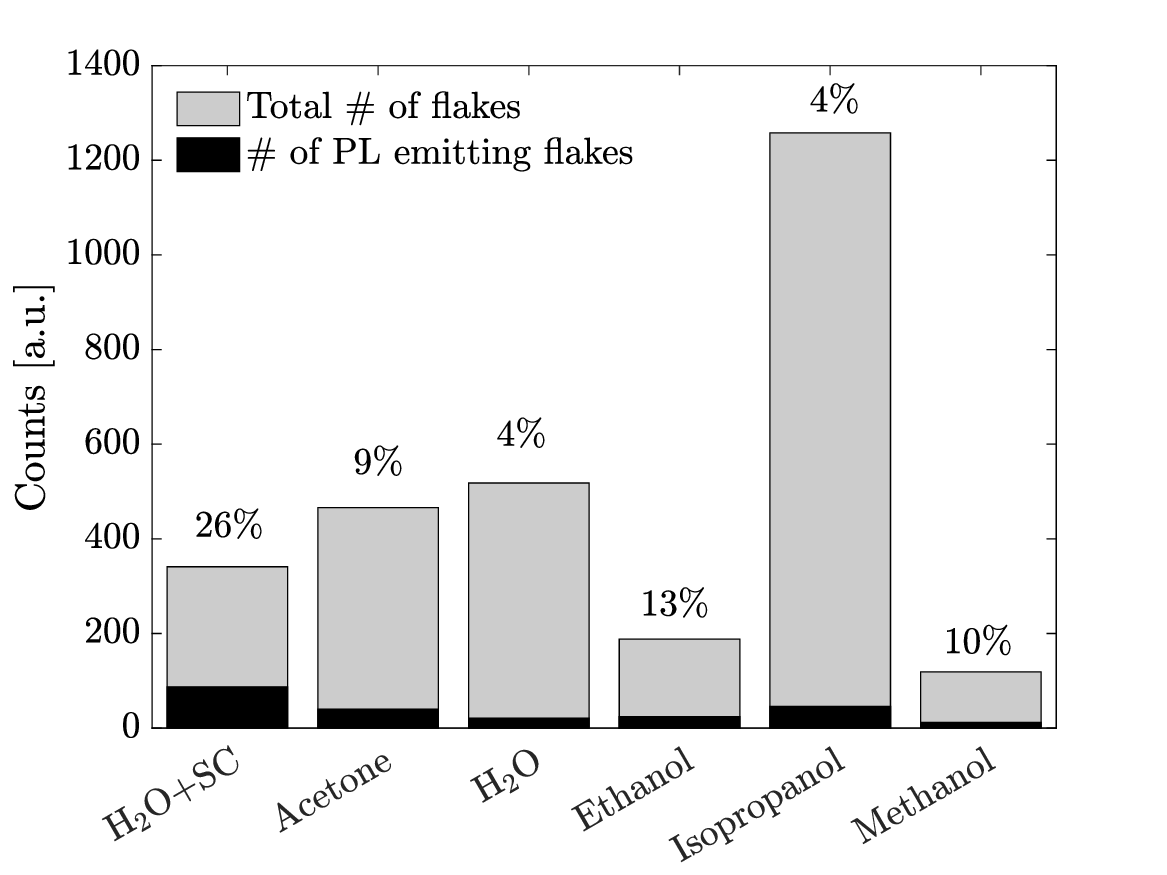}
    \caption{
    {\bf Ratio of photoluminescent (PL) flakes over the total number of deposited flakes on the substrate.}
    Comparison of all measured solvents for Merck's hBN show highest percentages for water and surfactant (SC) whereas water (H$_2$O) and isopropanol have the fewest amount of PL emitting flakes.
    }
    \label{figS4}
\end{figure*}

\clearpage

\begin{figure*}
    \includegraphics[width=\textwidth]{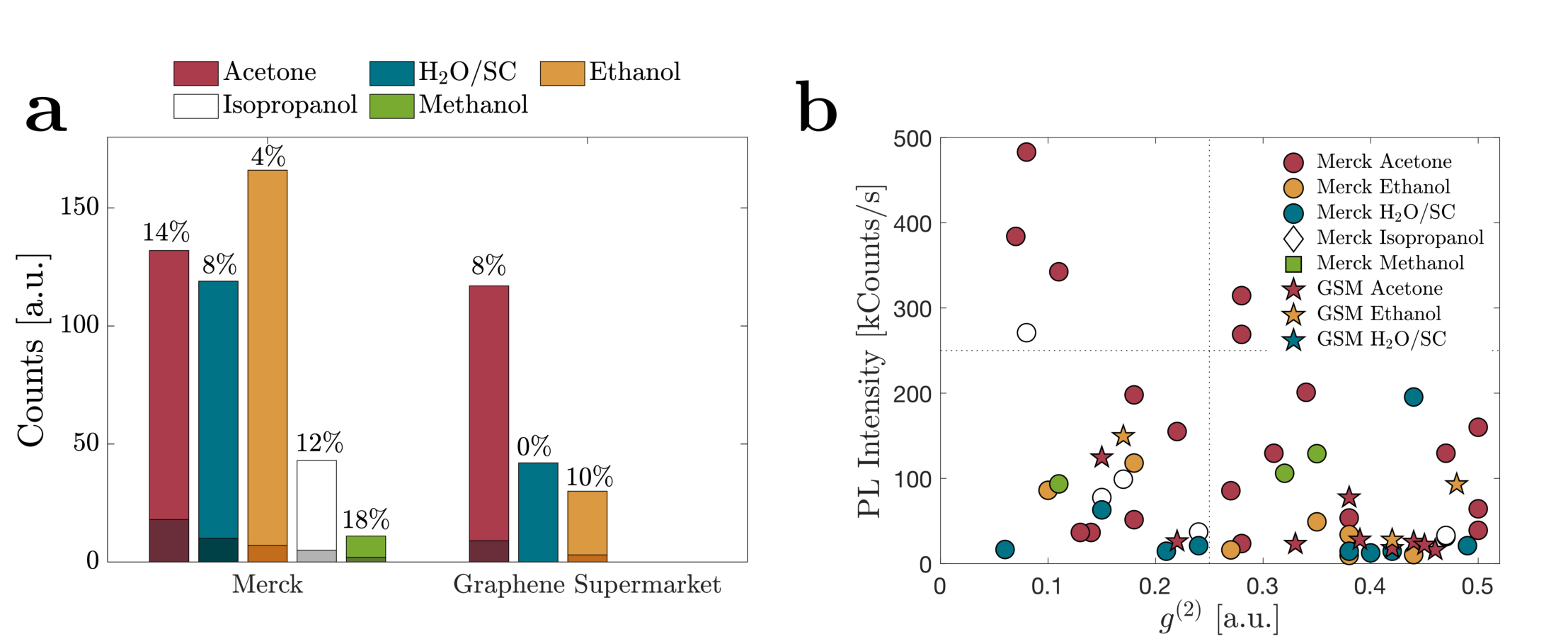}
    \caption{
    {\bf SPE yield and quality characterization for all measured solvents.}
    {\bf a} SPE ratio of all emitting flakes with additional data on isopropanol and methanol show that both possess also a high SPE ratio. {\bf b} Measured SPE in isopropanol and methanol are mostly found among the darker emitters. 
    }
    \label{figS5}
\end{figure*}
\clearpage

\begin{figure*}
    \includegraphics[width=.8\textwidth]{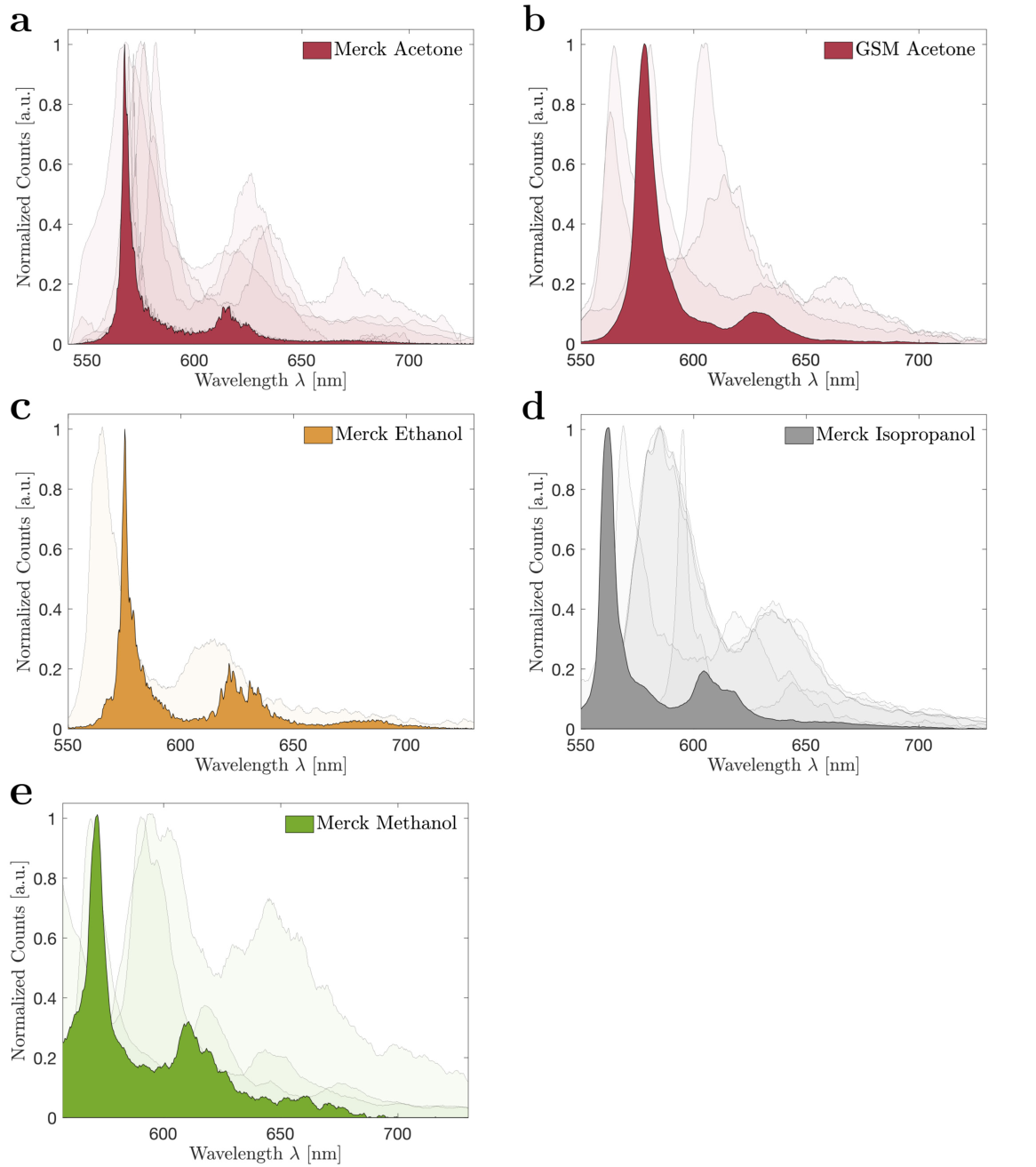}
    \caption{
    {\bf Multiple recordings of spectra within each solvent}
    {\bf a} Spectra of multiple single photon emitters in acetone for Merck's hBN, and {\bf b} for hBN from Graphene Supermarket (GSM). Remaining spectra are for Merck's hBN in {\bf c} ethanol, {\bf d} isopropanol, and {\bf e} methanol. All spectra show a similar variation of peak emissions between 550 and 600\,nm.
    }
    \label{figS6}
\end{figure*}

\clearpage
\begin{figure*}
    \includegraphics[width=\textwidth]{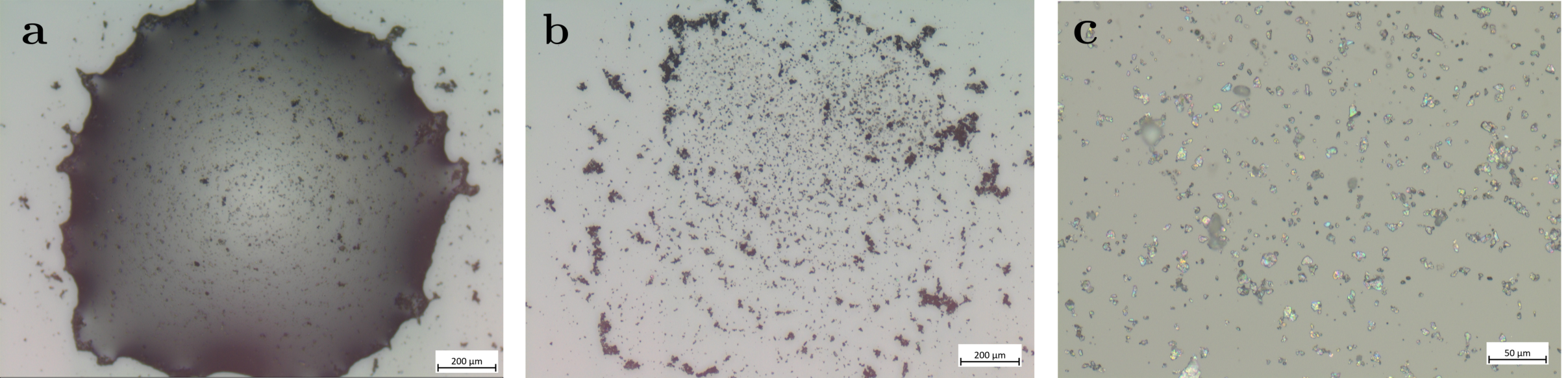}
    \caption{
    {\bf Drop-casting of hBN microflakes}
    Using micron-sized flakes of 2D Semiconductor (PWD-HBN), {\bf a} we take an  image directly after drop-casting with droplet still visible. {\bf b} Deposited flakes after drying of the droplet show large clusters accumulating with sizes above 10\,$\mu$m. {\bf c} Zoomed in image of the droplet in the center shows large cluster of flakes remaining on the substrate.
    }
    \label{figS7}
\end{figure*}

\begin{figure*}
    \includegraphics[width=.7\textwidth]{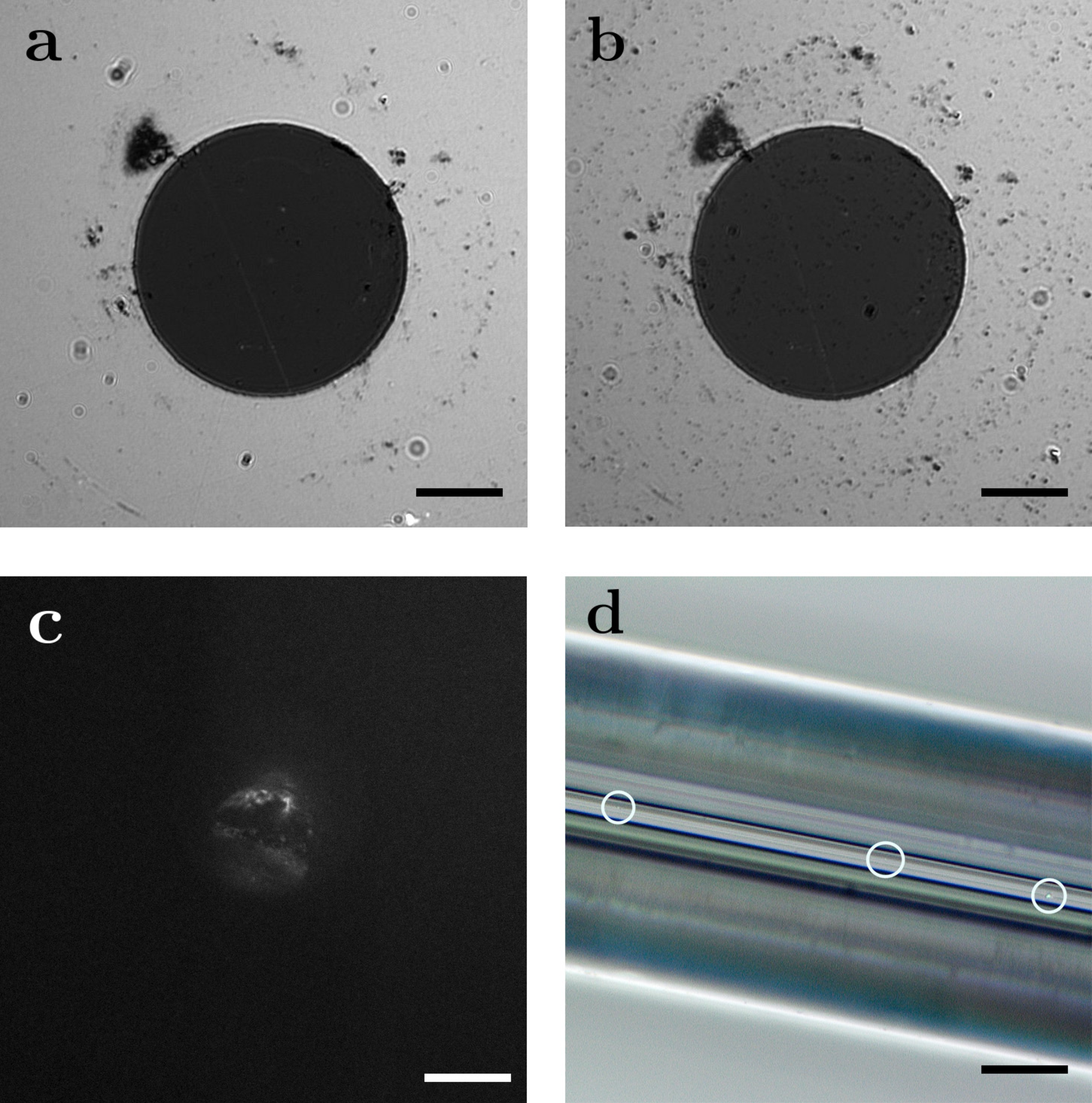}
    \caption{
    {\bf Applications of drop-casting on waveguides}
    {\bf a} The core of a single mode fiber under a 40x objective ($\rm NA=0.7$) can be {\bf b} dipped into a suspension of hBN nanoflakes leaving a droplet on its tip. After drying, individual nanoflakes are deposited on the fiber surface (black dots). {\bf c} Under a widefield PL microscope illumination from the other end of the fiber excites individual flakes visible as bright light source here. {\bf d} Even unevenly shaped waveguides such as exposed core fibers can be deposited with individual flakes (white circles) directly on its core without requiring complex alignment equipment. Scalebars {\bf a,b} $50\,\mu$m, {\bf b,d} $100\,\mu$m.
    }
    \label{figS8}
\end{figure*}

\clearpage

\begin{figure*}
    \includegraphics[width=\textwidth]{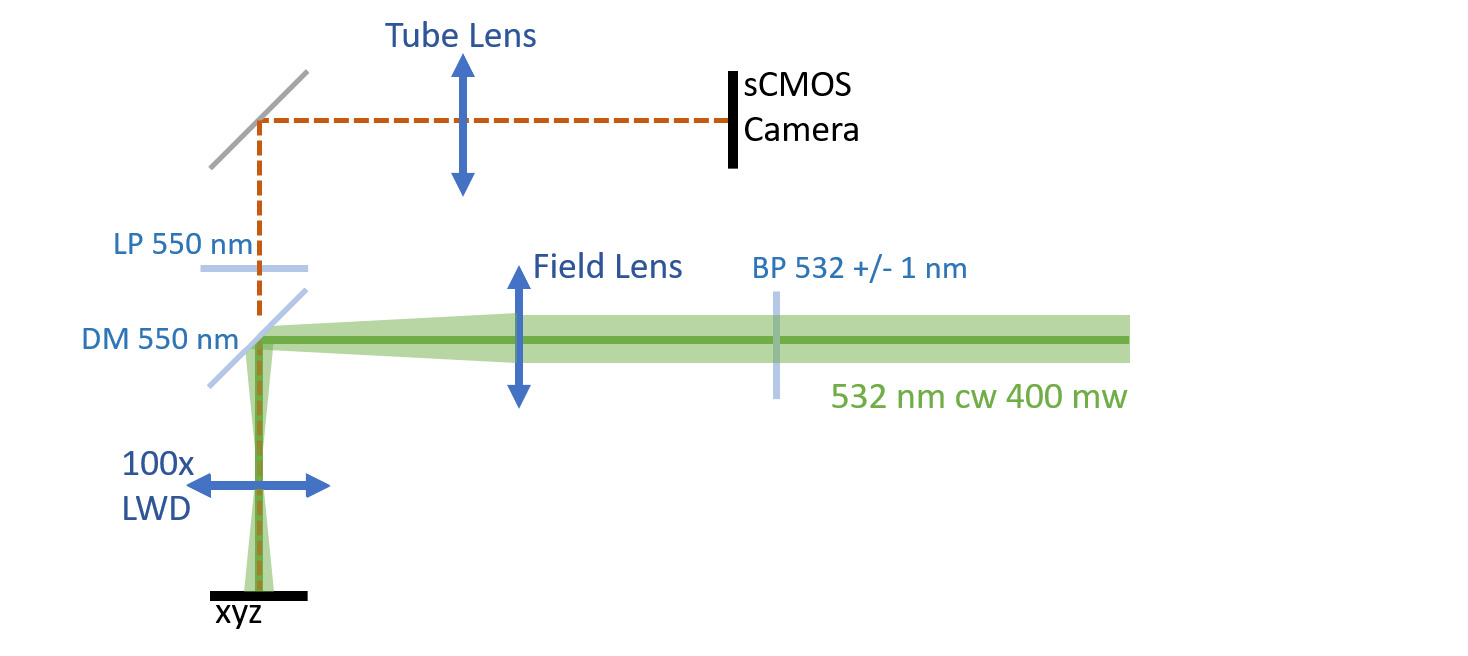}
    \caption{
    {\bf Experimental setup for wide-field photoluminescence microscopy}
    A light beam from a CW 532\,nm laser is focused with a field lens onto the back focal plane of a 100$\times$ long working distance (LWD) microscope objective.
    A longpass dichroic mirror (DM) and a long-pass filter only let the signal from the sample pass via a tube lens to a sCMOS camera. The sample's position can be controlled via x-y-z micrometer stage.
    }
    \label{figS9}
\end{figure*}

\clearpage
\begin{figure*}
    \includegraphics[width=.7\textwidth]{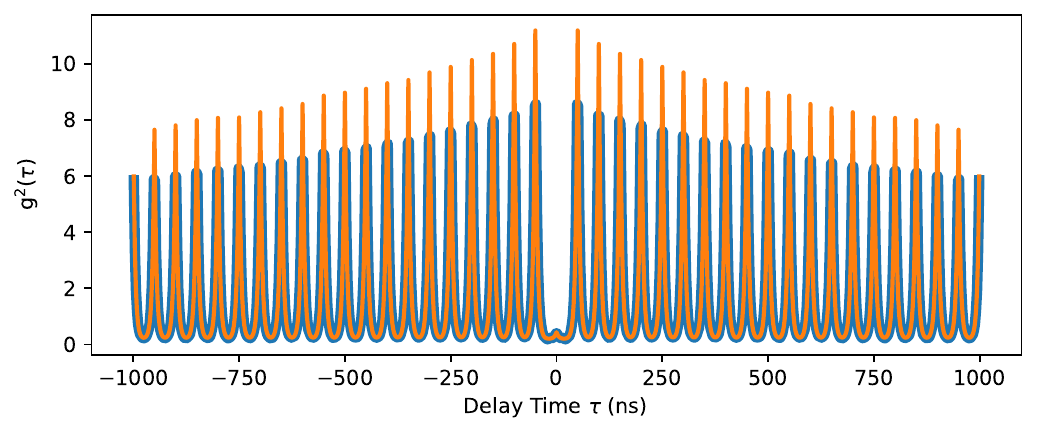}
    \caption{
    {\bf Life-time analysis of SPE} 
    Recorded $g^{(2)}$ correlation (blue line) over delay time $\tau$ for hBN from Merck and drop-casted in Acetone with fitted function (orange line) according to Eqn.~1.
    }
    \label{figS10}
\end{figure*}

\begin{figure*}
    \includegraphics[width=\textwidth]{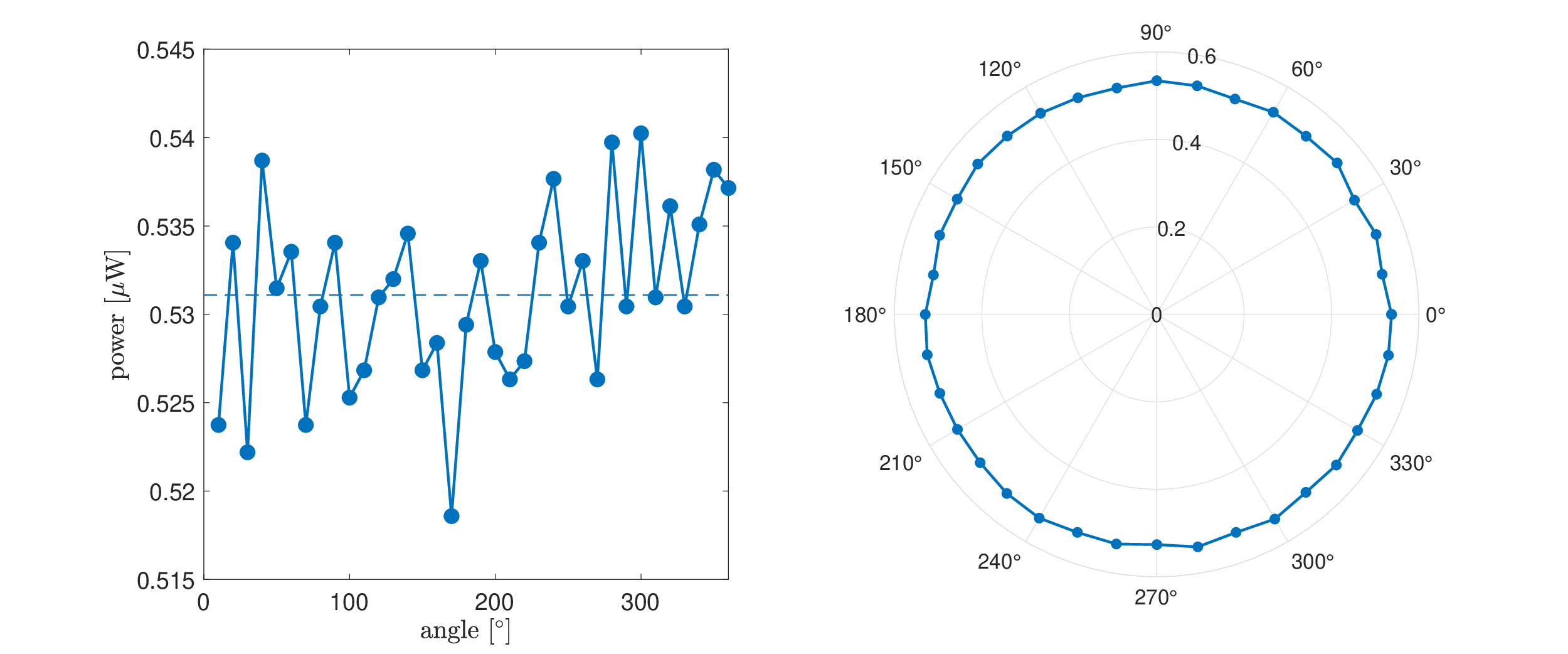}
    \caption{
    {\bf Polarization dependence of laser power} 
    The laser power remained constant over all polarization angles shown in a cartesian (left) and polar plot (right).
    }
    \label{figS11}
\end{figure*}

% Bibliography

\end{document}